\documentclass[conference]{IEEEtran}
\IEEEoverridecommandlockouts
\usepackage{cite}
\usepackage{amsmath,amssymb,amsfonts}
\usepackage{algorithmic}
\usepackage{multirow}
\usepackage{graphicx}
\usepackage{booktabs}
\usepackage{textcomp}
\usepackage{makecell}
\usepackage{array} 
\usepackage{xcolor}
\usepackage{balance}
\def\BibTeX{{\rm B\kern-.05em{\sc i\kern-.025em b}\kern-.08em
    T\kern-.1667em\lower.7ex\hbox{E}\kern-.125emX}}
\begin{document}

\title{LD-RPMNet: Near-Sensor Diagnosis for Railway Point Machines}

\author{
    \IEEEauthorblockN{
        Wei Li\IEEEauthorrefmark{1},
        Xiaochun Wu\IEEEauthorrefmark{1},
        Xiaoxi Hu\IEEEauthorrefmark{2},
        Yuxuan Zhang\IEEEauthorrefmark{3}, 
        Sebastian Bader\IEEEauthorrefmark{3},
        Yuhan Huang\IEEEauthorrefmark{4}
    }
    \IEEEauthorblockA{
        \IEEEauthorrefmark{1}\textit{School of Automation and Electrical Engineering, Lanzhou Jiaotong University, Lanzhou, China} \\
        \IEEEauthorrefmark{2}\textit{State Key Laboratory of Advanced Rail Autonomous Operation, Beijing Jiaotong University,  Beijing, China} \\
        \IEEEauthorrefmark{3}\textit{Department of Computer and Electrical Engineering, Mid Sweden University, Sundsvall, Sweden} \\  
        \IEEEauthorrefmark{4}\textit{School of Mechatronics Engineering, Harbin Institute of Technology, Harbin, China} \\ 
        Email:
        wxc@lzjtu.edu.cn
    }
}
%We Li 11230413@stu.lzjtu.edu.cn 
%Xiaochun Wu wxc@lzjtu.edu.cn
%Xiaoxi Hu xiaoxhu@bjtu.edu.cn
%Yuxuan Zhang yuxuan.zhang@miun.se
%Sebastian Bader sebastian.bader@miun.se
%Yuhan Huang 17771470797@163.com
%

\maketitle

\begin{abstract}
Near-sensor diagnosis has become increasingly prevalent in industry. This study proposes a lightweight model named LD-RPMNet that integrates Transformers and Convolutional Neural Networks, leveraging both local and global feature extraction to optimize computational efficiency for a practical railway application. The LD-RPMNet introduces a Multi-scale Depthwise Separable Convolution (MDSC) module, which decomposes cross-channel convolutions into pointwise and depthwise convolutions while employing multi-scale kernels to enhance feature extraction. Meanwhile, a Broadcast Self-Attention (BSA) mechanism is incorporated to simplify complex matrix multiplications and improve computational efficiency. Experimental results based on collected sound signals during the operation of railway point machines demonstrate that the optimized model reduces parameter count and computational complexity by 50\% while improving diagnostic accuracy by nearly 3\%, ultimately achieving an accuracy of 98.86\%. This demonstrates the possibility of near-sensor fault diagnosis applications in railway point machines.
\end{abstract}


\section{Introduction}
In recent years, significant progress has been made in intelligent operation and maintenance for various industrial fields~\cite{chen2025nonlinear,chen2025tensor}, especially the railway industry~\cite{9660775, liu2025accurate}. A Railway Point Machine (RPM) is an electromechanical device that moves and locks railway turnouts to ensure safe track switching. It mainly consists of an electric motor, a transmission system, and a locking mechanism working together to control and secure the turnout position. Due to its vital role in railway operations, accurate fault diagnosis of RPMs has become a key research focus, especially with the rapid advancement of artificial intelligence technologies~\cite{huxx,xs,chenxh}. 

Existing studies primarily focused on complex fault detection and diagnosis methods, with limited attention given to the challenges of computational efficiency and real-time application in practical engineering scenarios~\cite{hu2, adin2023b, huang2024b}.Hu et al.~\cite{hu1} employed vibration signals collected from multiple locations on the machine to enhance diagnostic accuracy. Liu et al.~\cite{liu2023modified, liu2024multi} used collected sound signals, and combined feature engineering with machine learning to achieve fault diagnosis of the RPM.

Nevertheless, these studies have neglected the high computational cost and real-time constraints of the proposed models, which hinders their deployment and widespread adoption in real-world applications~\cite{hu3, zhangtim2023a, Ud2024}. In contrast, many near-sensor and in-sensor computations have been explored in the field of structural health monitoring and in the field of medical health monitoring. For example, Zhang et al. compared lightweight Convolutional Neural Networks (CNNs) for crack detection~\cite{zsas2024b,Y2025i2mtc,zlcn2022}. Giordano et al.~\cite{10599159} designed a lightweight machine learning algorithm for digital biomarker-based sepsis alerts. 

Inspired by these studies and considering the constraints on cost and diagnostic accuracy in the railway industry, we establish a Lightweight Diagnosis of Railway Point Machines Network (LD-RPMNet) based on sound signals when the RPM operates. To fully exploit the rich spectral information embedded in sound signals, we employ multi-scale convolution for effective time-domain feature extraction. Meanwhile, we introduce the Transformer architecture to model global temporal dependencies, thereby improving the capability of capturing long-term features. Importantly, to meet the requirements of practical near-sensor applications, we optimize both core modules for lightweight implementation, reducing computational overhead and enhancing real-time performance, thus enabling efficient fault diagnosis in near-sensor environments.

\section{Methodology and Scheme}
\begin{figure*}[t]
	\centering
	\includegraphics[width=0.6\textwidth]{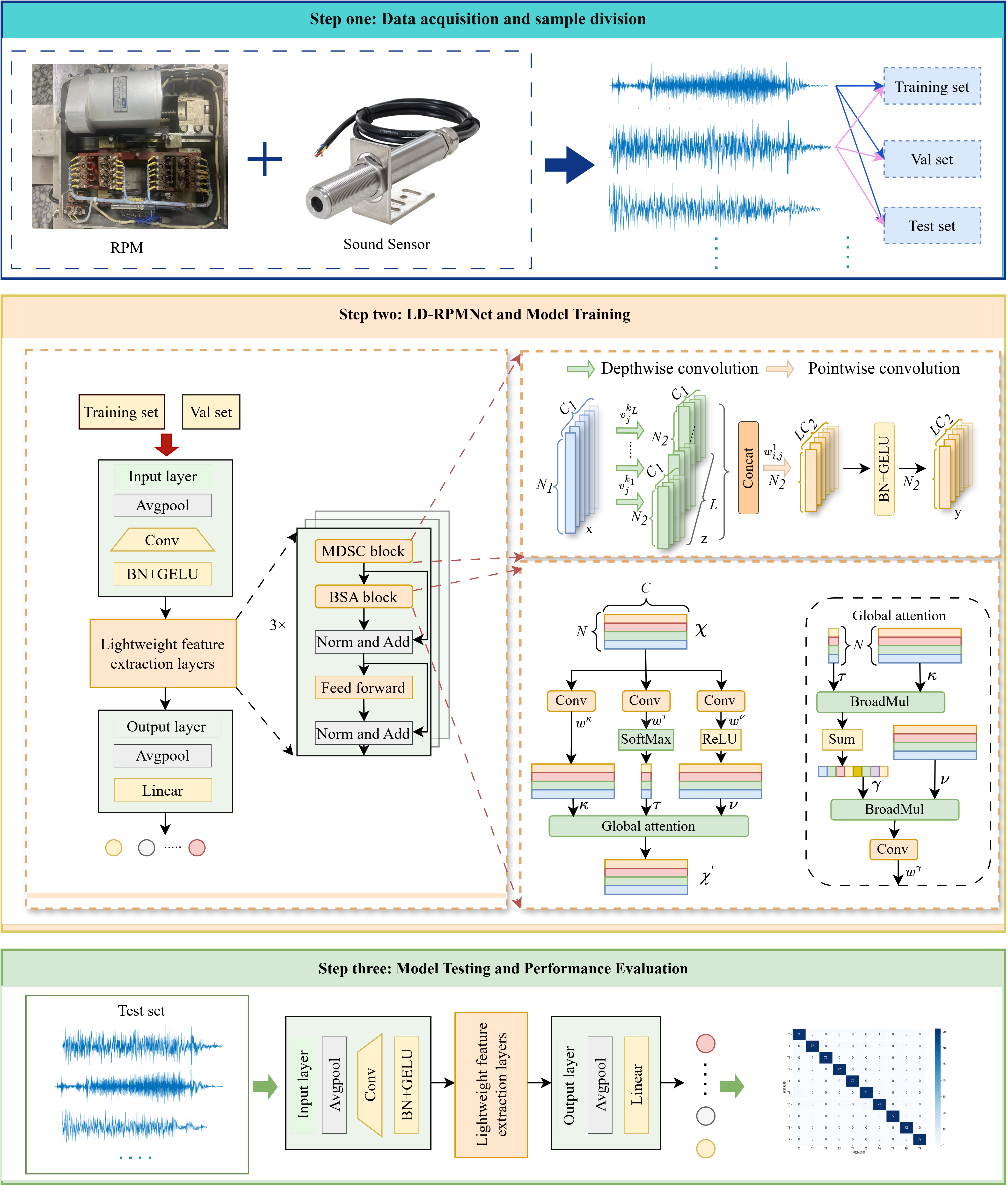}
	\caption{Overview of the proposed near-sensor RPM fault diagnosis approach and its implementation.}
	\label{fig6}
\end{figure*}

The framework (illustrated in Fig.~\ref{fig6}) of our approach and its implementation can be structured into three key stages: (1) data collection and sample division, (2) LD-RPMNet and model training, (3) model testing and performance evaluation.

\begin{figure}[t]
        \vspace{-8mm}
	\centering
	\includegraphics[width=0.35\textwidth]{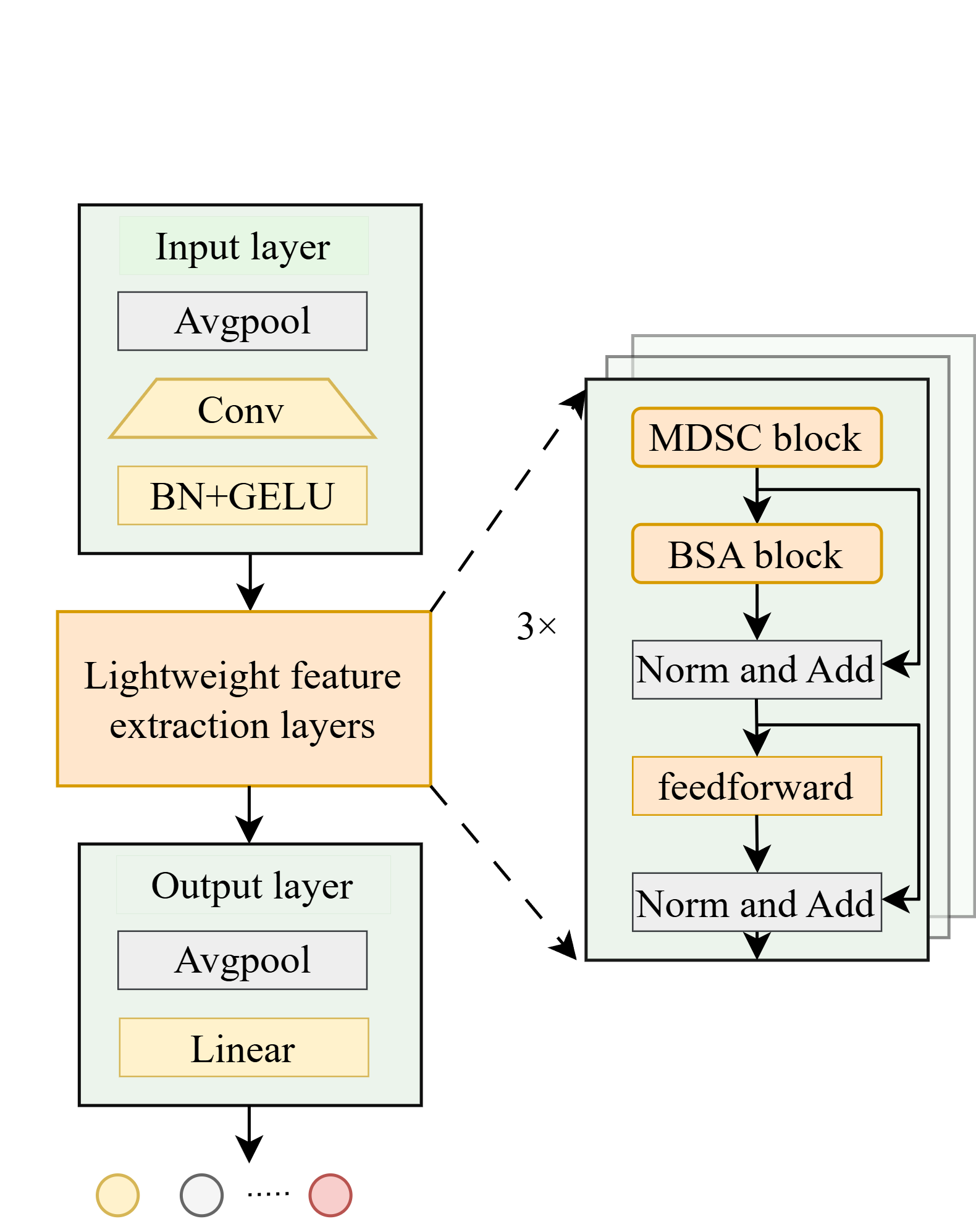}
	\caption{The network diagram of the LD-RPMNet.}
	\label{fig7}
\end{figure}

For near-sensor diagnosis, we propose LD-RPMNet to enhance data processing efficiency, while reducing computational complexity and resource consumption, as shown in Fig.~\ref{fig7}. It comprises two lightweight modules: Multi-scale Depthwise Separable Convolution (MDSC) and Broadcast Self-Attention (BSA). MDSC enhances feature extraction by decomposing cross-channel convolutions into pointwise and depthwise operations with multi-scale kernels, making it ideal for low-power, resource-constrained environments. Meanwhile, BSA simplifies matrix operations, reducing computational overhead. This design minimizes reliance on cloud computing, lowers bandwidth requirements, and improves data privacy. By integrating MDSC and BSA, LD-RPMNet aims to optimize the near-sensor diagnosis pipeline, enhancing real-time model performance.

\subsection{Multi-Scale Depthwise Separable Convolution}\label{sec21}
CNN employs cross-channel convolutions, which require weighted summation across all input channels to integrate channel information, resulting in a large number of parameters and high computational complexity. Moreover, the use of single-scale convolutional kernels limits the scale of feature extraction. Therefore, lightweight multi-scale feature extraction is crucial for improving efficiency and performance.

This paper proposes MDSC, an innovative module that combines multi-scale convolutional kernels with depthwise separable convolution to achieve a lightweight optimization of CNNs, making it more suitable for near-sensor diagnosis. As illustrated in Fig.~\ref{fig8}, MDSC provides the following two primary optimizations:

\begin{figure}[t]
	\centering
	\includegraphics[width=0.45\textwidth]{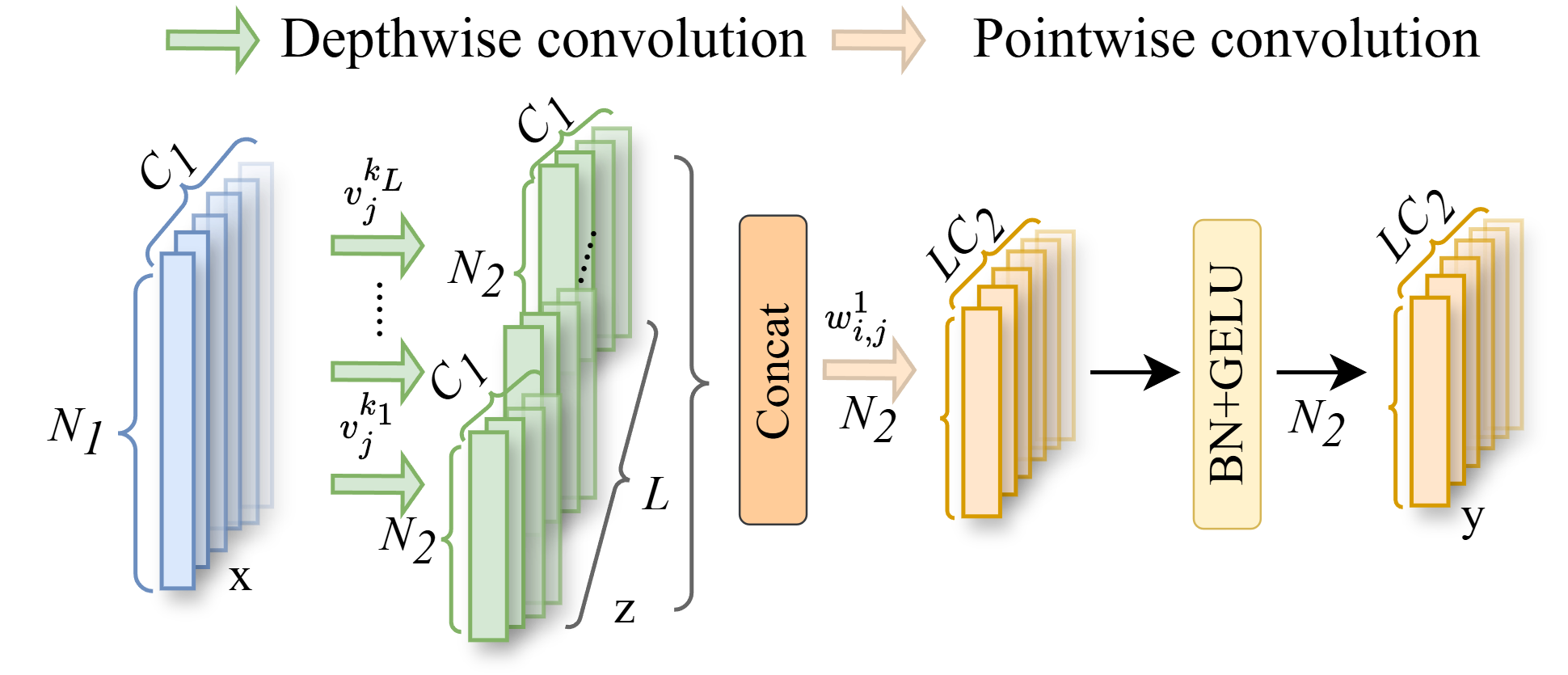}
	\caption{The structure of MDSC.}
	\label{fig8}
\end{figure}

\textbf{(1) Cross-Channel Convolution Optimization:} MDSC optimizes cross-channel convolution by dividing it into depthwise convolution and pointwise convolution. Depthwise convolution applies convolution operations independently to each input channel, while pointwise convolution utilizes a convolution with a kernel size of 1 to operate across all channels. This method considerably decreases the quantity of parameters and computational intricacy.

\textbf{(2) Multi-Scale Convolutional Kernels:} Multiple scales of convolutional kernels are utilized to simultaneously extract features at different scales. These kernels operate in parallel to extract features from various local scales, enabling the capture of detailed information across multiple scales while also reducing computational load.

Initially, convolutional kernels of varying sizes are applied in parallel. The depthwise convolution ensures that each output channel corresponds to a specific input channel, effectively reducing the number of learnable parameters and computational time while retaining the ability to capture information from multiple local receptive fields according to

\begin{equation}
Z^{k_l} = \mathrm{Concat}_{j=1}^{C_1}\left( \mathbf{v}_j^{k_l} \times \mathbf{x}_j \right),
\end{equation}
where $\mathbf{x}_j \in \mathbb{R}^{C_1 \times N_1}$ represents the input; 
$\mathbf{z} \in \mathbb{R}^{C_1 \times N_2}$ represents the data obtained after depthwise convolution; 
$\mathbf{v}^{k_l} \in \mathbb{R}^{C_2 \times k_l}$ denotes the convolution weights for a convolution kernel of size $k_l$; 
$\times$ represents the convolution operation; 
$\mathrm{Concat}(\cdot)$ represents the concatenation operation; 
$C_1$ represents the input channel dimension,
$N_1$ denotes the input time dimension,
$C_2$ refers to the output channel dimension, and
$N_2$ indicates the output time dimension.

Subsequently, the output signals from the depthwise convolution are concatenated, and a cross-channel convolution with a kernel size of 1 is employed to integrate information across channel dimensions and adjust the dimensionality of the channels. This process is expressed as:

\begin{equation}
y' = \mathrm{Concat}_{j=1}^{LC_2} \left( \sum_{i=1}^{LC_1} \mathbf{w}_{i,j}^1 \times \mathbf{z}_i \right),
\end{equation}
where $\mathbf{w}_{i,j}^1 \in \mathbb{R}^{C_2 \times C_1}$ represents the convolution weights for a kernel size of 1; $L$ represents the number of different convolution kernels.

Finally, the local features are aggregated and passed through a Batch Normalization layer, followed by the application of a Gaussian Error Linear Unit (GELU) activation function, resulting in the final feature representation.

\subsection{Broadcast Self-Attention}
\begin{figure}[t]
	\centering
	\includegraphics[width=0.45\textwidth]{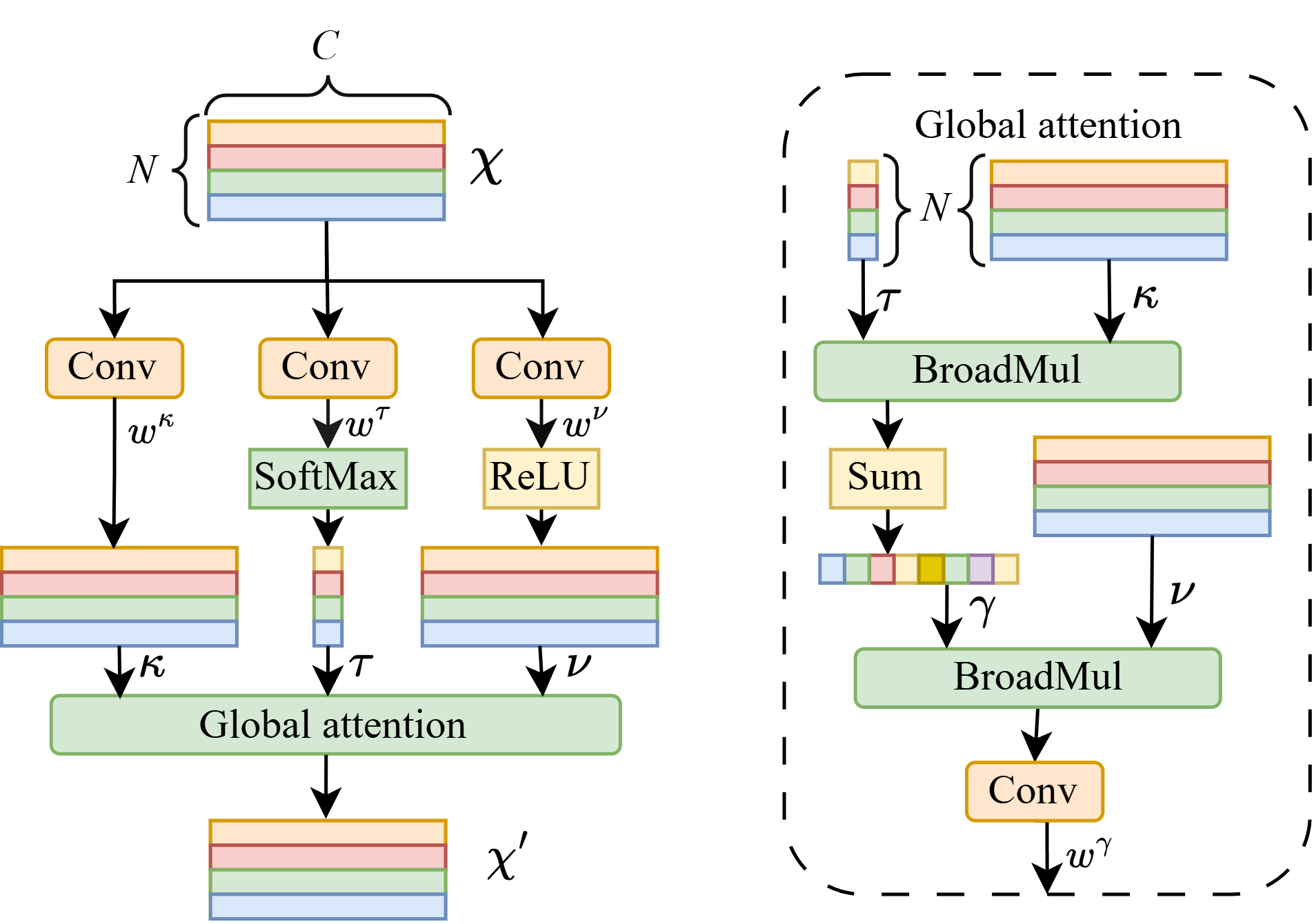}
	\caption{The structure of BSA.}
	\label{fig9}
\end{figure}

Multi-Head Self-Attention (MHSA) is a crucial component of the Transformer architecture~\cite{he2023msit,3,he2024cfspt}. Initially, linear transformations are applied to generate the query, key, and value matrices. For each attention head, the dot product between the query and key matrices is computed, followed by the application of the softmax function to derive the attention weights. These weights are then used to perform a weighted sum of the value matrix, yielding the output for each head. The outputs from multiple heads are subsequently concatenated and linearly transformed to produce the final output representation.

However, the matrix multiplications and multi-dimensional exponential operations involved in MSHA result in high computational complexity, particularly when handling high-dimensional data. This significantly increases the computational cost and hardware requirements of the model. To address this issue, the BSA mechanism was introduced~\cite{25}. By leveraging broadcast operations, BSA avoids complex matrix multiplications and multi-dimensional exponential operations. This improvement not only reduces the number of parameters and computational load but also enhances the model's performance in resource-constrained environments.

This paper introduces BSA to achieve a lightweight optimization of MSHA, as illustrated in Fig.~\ref{fig9}. The proposed method is particularly suitable for near-sensor diagnosis. By leveraging broadcast operations, BSA provides optimizations in the following two key aspects:

\textbf{(1) Reducing Parameters:} BSA minimizes the number of parameters and computational load by avoiding matrix multiplication and multi-dimensional exponential operations through broadcast operations. This optimization significantly reduces model complexity and computational resource requirements while preserving efficient feature extraction capabilities.

\textbf{(2) Lightweight Feature Extraction:} BSA effectively integrates and propagates critical feature weight information across the global scope of the signal through broadcast operations, thereby enhancing the comprehensiveness and accuracy of feature extraction.

\section{Dataset and Experiment Settings}
\begin{figure}[t]
	\centering
	\includegraphics[width=0.45\textwidth]{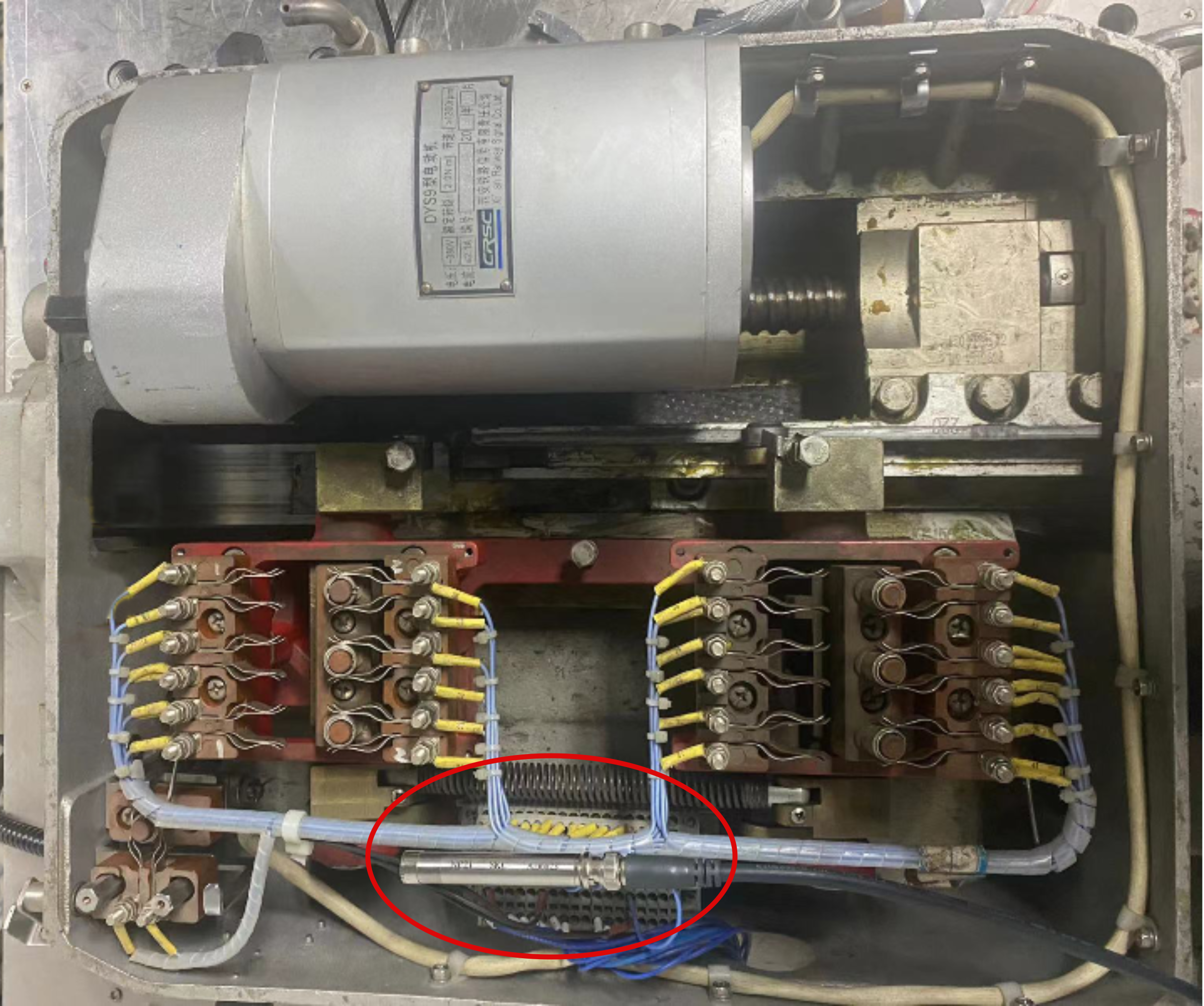}
	\caption{ZDJ9-type RPM and sound sensor location.}
	\label{fig10}
\end{figure}

\begin{table}[t]
\centering
\caption{Sample distribution and fault phenomena}
\label{tab:fault_phenomena}
\begin{tabular}{>{\centering\arraybackslash}m{1cm}>{\centering\arraybackslash}m{1cm}>{\raggedright\arraybackslash}m{5cm}} \toprule
\textbf{Class} & \textbf{Sample count} & \textbf{Fault phenomenon} \\ \midrule
1  & 121  & Normal (4kN) \\
2  & 180  & Abnormal energy supply \\
3  & 158  & Unstable power supply \\
4  & 120  & Underdriving (3kN) \\
5  & 124  & Overdriving (5kN) \\
6  &  84  & Overdriving (6kN) \\
7  &  80  & Overdriving (8kN) \\
8  & 113  & Sudden indication loss after approaching movement \\
9  & 112  & Unlocking failure \\
10 & 120  & No-load \\ \bottomrule
\end{tabular}
\end{table}
\subsection{Dataset}

This study employs sound signals from the switching process of a ZDJ9-type RPM as the data source for model training. The placement of the sound sensors is illustrated in Fig.~\ref{fig10}. Considering the high sensitivity, low noise, and excellent reliability of the NP21 IEPE preamplifier in industrial environments, this device was selected for data acquisition under clear weather conditions without additional noise interference. The dataset comprises 10 classes, including the normal operational state of the ZDJ9-type RPM and 9 typical fault types, as detailed in Table~\ref{tab:fault_phenomena}.

\begin{figure}[t]
	\centering
	\includegraphics[width=0.48\textwidth]{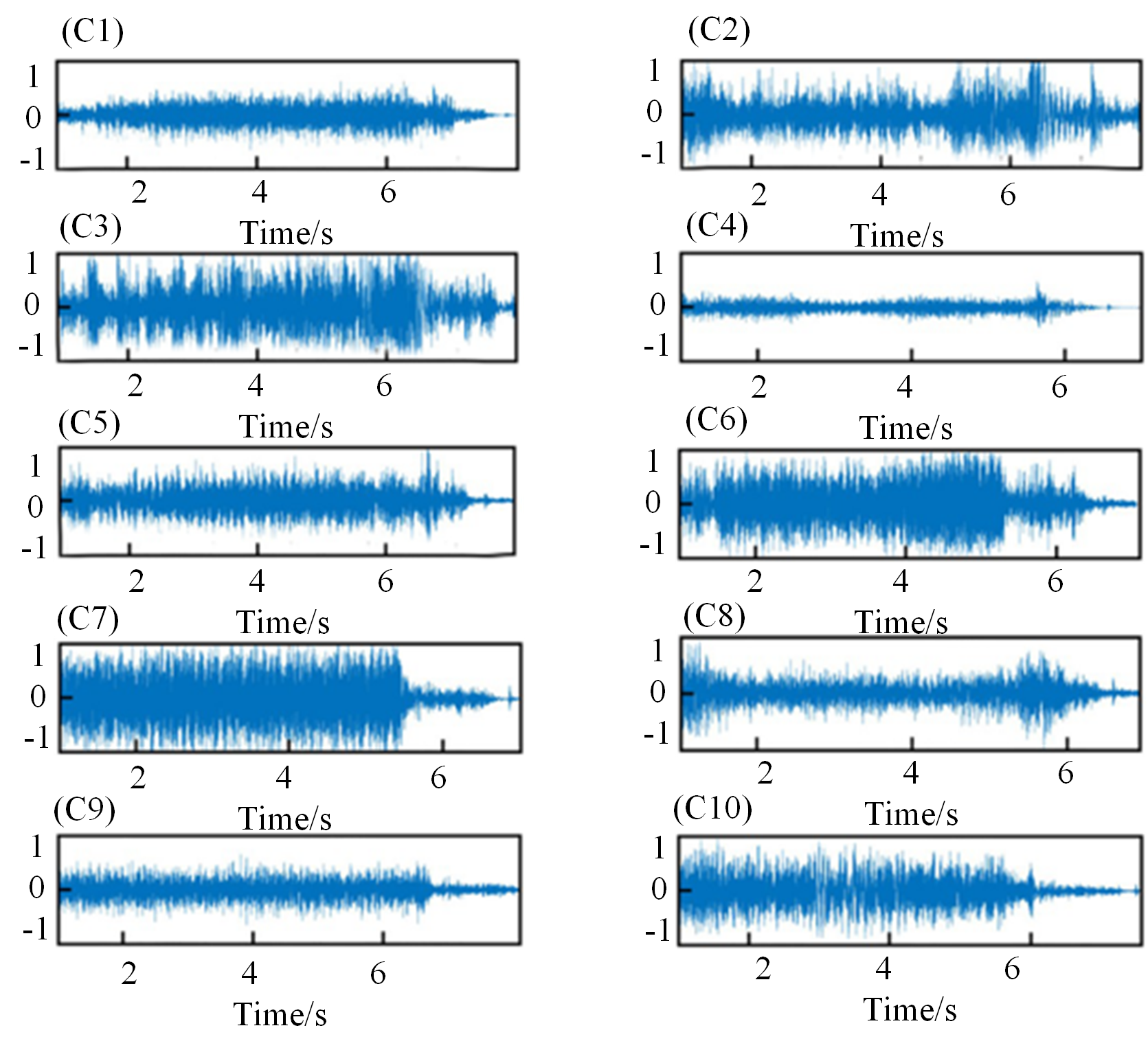}
	\caption{Sound signal waveforms for various faults of the ZDJ9-type RPM.}
	\label{fig13}
\end{figure}

Examples of sound signals for the different operational states are presented in Fig.~\ref{fig13}. Fault (C2) is typically caused by an issue in one phase of the three-phase power supply, whereas fault (C3) is usually due to unstable terminal contact. Although the underlying causes of faults (C2) and (C3) differ, their fault waveform characteristics are similar, making them difficult to distinguish. Fault (C3) exhibits smaller signal amplitudes due to an under-driven waveform, while faults (C5) through (C7) demonstrate progressively increasing vibrations caused by overload waveforms.

\subsection{Experiment Settings}
The experiments were conducted on a Windows 10 system equipped with an Intel Core i5-12700K CPU, 16 GB of RAM, and an NVIDIA RTX 4060 GPU with 12 GB of VRAM. The fault diagnosis classification network was implemented using PyTorch 2.0, and both the training and inference processes were accelerated using the GPU.

The 1212 sound signals were split into training, validation, and test sets (7:2:1 ratio), yielding 845, 245, and 122 samples, respectively. The training and validation sets were used for model training, while the test set evaluated performance. The AdamW optimizer and cross-entropy loss function were applied for multi-class classification. Balancing speed and accuracy, the batch size and learning rate were set to 16 and 0.001. With convergence observed at 40 epochs, training was conducted for 50 epochs.

\section{Experimental Results and Analysis}

\subsection{Ablation Study}

This section conducts ablation experiments using the sound dataset to evaluate the effectiveness and necessity of MDSC and BSA in the context of near-sensor applications. Here, CNT represents the baseline CNN+transformer model. Four methods are evaluated: the first is the CNT algorithm, the second is the CNT+MDSC algorithm, the third is the CNT+BSA algorithm, and the fourth is the LD-RPMNet algorithm, as detailed in Table~\ref{tab3}.

\begin{figure}[t]
	\centering
	\includegraphics[width=0.48\textwidth]{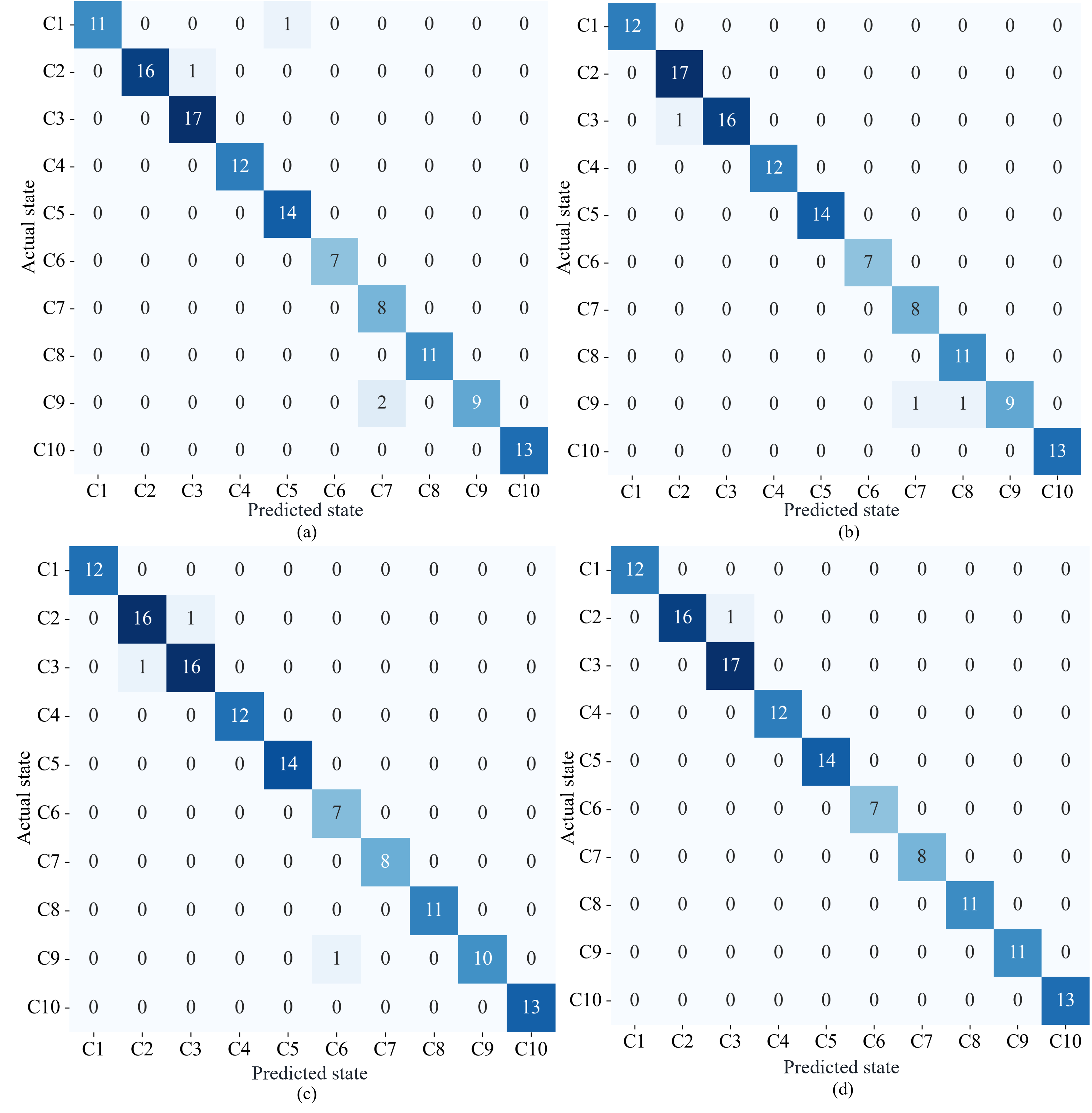}
	\caption{Confusion matrix of ablation study: 
 (a) CNT, (b) CNT+MDSC, (c) CNT+BSA, (d) LD-RPMNet.}
	\label{fig11}
\end{figure}

\begin{table}[t]
    \centering
    \caption{Reaults of ablation study (Accuracy in \%, FLOPs in M)}
    \label{tab3}
    \begin{tabular}{cccccc} \toprule
    \textbf{Method} & \textbf{CNT} & \textbf{MDSC} & \textbf{BSA} & \textbf{Accuracy} & \textbf{FLOPs} \\ \midrule
    1 & \checkmark & & & 96.60 & 39.41 \\
    2 & & \checkmark & & 97.12 & 28.91 \\
    3 & & & \checkmark & 97.42 & 31.48 \\
    4 & \checkmark & \checkmark & \checkmark & 98.86 & 20.96 \\ \bottomrule
    \end{tabular}
\end{table}

Table~\ref{tab3} and Fig.~\ref{fig11} present the results of the ablation experiments. The results demonstrate that the incorporation of the MDSC and BSA modules not only improves the model's accuracy but also reduces the parameter count and computational cost. These results validate the effectiveness of the MDSC and BSA modules in near-sensor diagnosis.

\subsection{Comparative Experiment}
This paper conducts a comparative experiment of six algorithms. The LD-RPMNet refers to the algorithm proposed in this work. $\textnormal{Convformer\_NSE}$~\cite{21}, and CLFormer~\cite{20} are lightweight CNN+transformer hybrid algorithms introduced in other studies. MobileNet~\cite{23} and ResNet18~\cite{24} are widely utilized CNN algorithms.

\begin{figure}[t]
	\centering
	\includegraphics[width=0.48\textwidth]{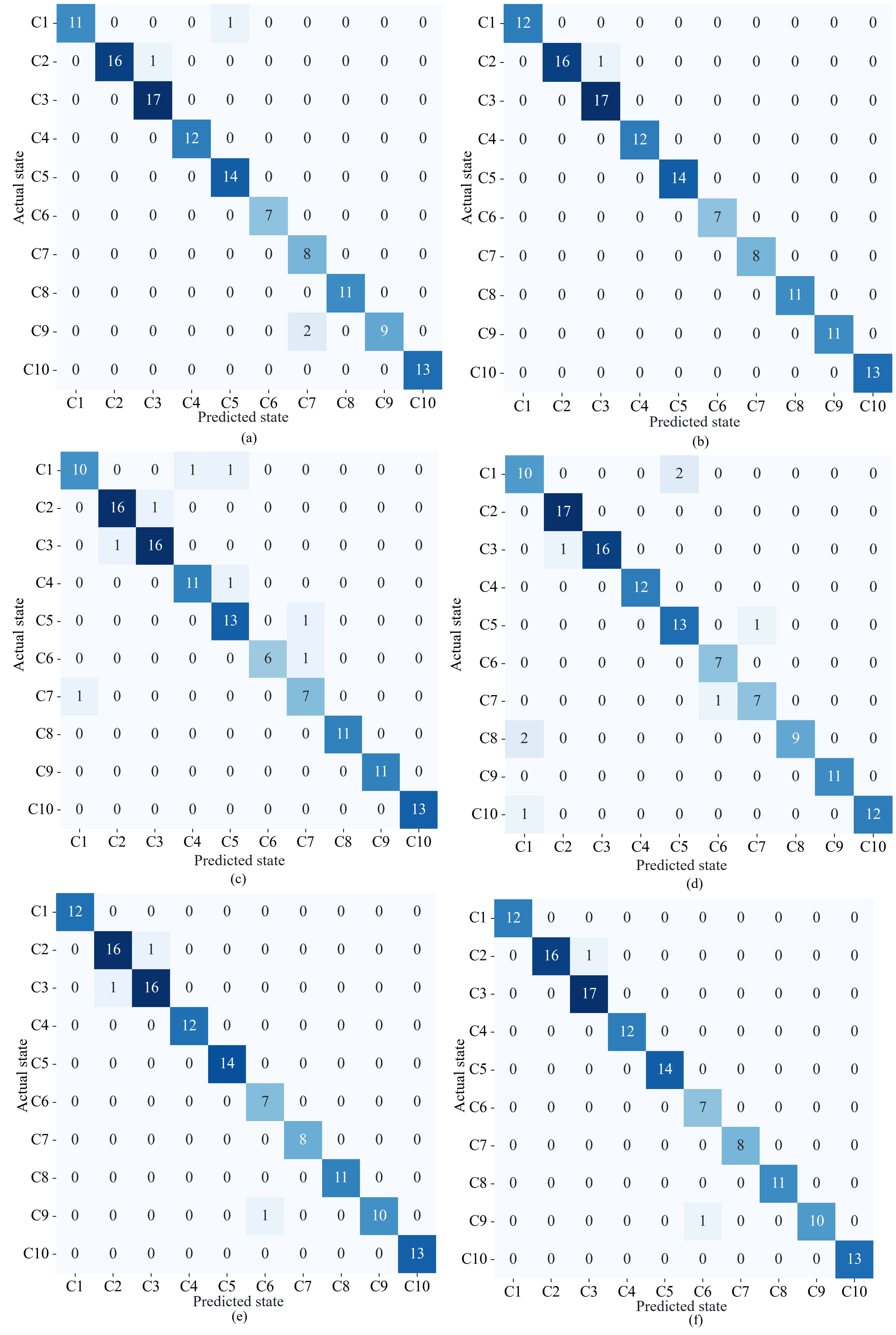}
	\caption{Confusion matrix of comparative experiment: (a) CNT, (b) LD-RPMNet, (c) CLFormer, (d) Convformer\_NSE, (e) MobileNet, (f) ResNet18.}
	\label{fig12}
\end{figure}

Comparative experiments were conducted on the test set results. The evaluation metrics in this study include accuracy, precision, recall, F1-score, number of parameters, computational cost, and inference time. The detailed results are presented in Table~\ref{tab:results} and Fig.~\ref{fig12}. 

\begin{table*}[t]
\centering
\caption{Performance comparison of different methods (Accuracy, Precision, Recall in \%, FLOPs, Params in M, Inference time in s)}
\begin{tabular}{lccccccc}
\hline
\textbf{Method} & \textbf{Accuracy} & \textbf{Precision} & \textbf{Recall} & \textbf{F1-score} & \textbf{Params} & \textbf{FLOPs} & \textbf{Inference Time} \\
\hline
CNT & 96.60 & 100 & 98.33 & 99.16 & 1.30 & 39.41 & 0.032 \\
{LD-RPMNet} &{98.86} & {100} &{100} & {100} & {0.48} & {20.96} & {0.025} \\
CLFormer & 90.30 & 99.83 & 93.33 & 96.47 & 0.01 & 0.133 & 0.021 \\
Convformer\_NSE & 95.08 & 99.83 & 95.01 & 97.36 & 0.24 & 6.22 & 0.038 \\
MobileNet & 97.50 & 100 & 100 & 100 & 3.19 & 333.52 & 0.110 \\
ResNet18 & 99.14 & 100 & 100 & 100 & 3.85 & 175.69 & 0.036 \\
\hline
\end{tabular}
\label{tab:results}
\end{table*}

Comparative experimental results indicate that LD-RPMNet demonstrates outstanding performance in classification accuracy, only slightly lower than that of ResNet18. LD-RPMNet also excels in precision and recall, indicating its strong capability in distinguishing between normal and faulty equipment states, which is crucial for fault diagnosis and maintenance in railway systems. Furthermore, LD-RPMNet maintains high accuracy while significantly reducing model size, with a parameter count that is over seven times smaller than that of ResNet18 and 50\% lower than that of the CNT algorithm. Although CLFormer and $\textnormal{Convformer\_NSE}$ have fewer parameters, their classification accuracy is reduced by 8\% and 4\%, respectively. In terms of computational complexity, LD-RPMNet significantly reduces computational costs, achieving an approximately eightfold reduction compared to ResNet18 and nearly 50\% compared to CNT. Additionally, it exhibits a shorter inference time, further enhancing computational efficiency. Overall, the experimental results confirm that LD-RPMNet offers substantial advantages in classification accuracy, parameter efficiency, and computational efficiency, reinforcing its effectiveness in near-sensor intelligent fault diagnosis.

\section{Conclusion}
This paper presents an innovative fault diagnosis model for RPMs, which significantly reduces the number of parameters and computational complexity, effectively addressing the requirement posed for near-sensor fault diagnosis. By enabling near-sensor diagnosis, the model not only minimizes data transmission requirements but also enhances real-time performance and data privacy, making it more suitable for resource-constrained environments. Future work plans to investigate the implementation of our method on real-time microcontrollers or edge systems to evaluate its performance in real-world applications.

\section*{Acknowledgement}
The authors would like to thank the financial support by the Knowledge Foundation under grant 20180170 (NIIT) and 20240029-H-02 (TransTech2Horizon).

\balance
\bibliographystyle{IEEEtran}
\bibliography{references}

\end{document}